# Active-Spin-State-Derived Descriptor for Hydrogen Evolution Reaction Catalysis

Yu Tan,[1] Lei Li,[1] Zi-Xuan Yang,[1] Tao Huang,[1] Qiao-Ling Wang,[2] Tao Zhang,[1] Jing-Chun Luo,[1] Gui-Fang Huang,[1]* Wangyu Hu,[3] Wei-Qing Huang[1]*

[1]*Department of Applied Physics, School of Physics and Electronics, Hunan University, Changsha 410082, China*
[2]*Changsha Environmental Protection College, Changsha 410004, China*
[3]*School of Materials Science and Engineering, Hunan University, Changsha 410082, China*

**Abstract:** Spin states are pivotal in modulating the electrocatalytic activity of transition-metal (TM)-based compounds, yet quantitatively evaluating the activity-spin state correlation remains a formidable challenge. Here, we propose an 'activity index ($\eta$)' as a descriptor, to assess the activity of the spin states for the hydrogen evolution reaction (HER). $\eta$ descriptor integrates three key electronic parameters: the proportion ($P$), broadening range ($R$) and center ($\varepsilon_{act}$) of active spin state, which collectively account for the electronic structure modulation induced by both the intrinsic active site and its local coordination environment. Using 1T-phase $ZrSe_2$-anchored TM atoms (TM=Sc to Ni) as prototypes, we reveal that the correlation between Gibbs free energy ($\Delta G_H$) and the $\eta$ value follows a linear relation, namely, the $\Delta G_H$ reduces as the $\eta$ decreases. Notably, $ZrSe_2$-Mn exhibits the optimal $\eta$ value (-0.56), corresponding the best HER activity with a $\Delta G_H$ of 0.04 eV— closer to the thermoneutral ideal value (0 eV) than even Pt ($\Delta G_H$ = -0.09 eV). This relationship suggests that $\eta$ is the effective descriptor of active spin state for HER of TM-based catalysts. Our study brings fundamental insights into the HER activity-spin state correlation, offering new strategies for HER catalyst design.

*. Corresponding authors: wqhuang@hnu.edu.cn, gfhuang@hnu.edu.cn

Electrochemical water splitting has emerged as a promising approach for efficient, green, and sustainable hydrogen production, yet its scalability is limited by the efficiency and cost of current catalysts.[1–5] While platinum remains the benchmark for the hydrogen evolution reaction (HER), its scarcity drives urgent efforts to develop earth-abundant alternatives.[6–11] Typically, the rational design of HER catalysts primarily focuses on tuning electronic properties, including valence states and $d$-band center positions.[12–18]

Recent experimental and theoretical investigations demonstrate the critical role of spin states in transition metal (TM)-based catalysts, offering an underexplored avenue to tailor catalytic activity beyond traditional electronic structure modulation.[19–23] In TM-based single-atom catalysts (SACs), high-spin (HS) and low-spin (LS) configurations exhibit distinct interactions with reaction intermediates, enabling spin-dependent reactivity. For instance, for oxygen evolution reaction (OER) process, $Co^{3+}$ in HS state facilitates stronger $d$-$p$ orbital overlap with oxygen-containing species, lower the energy barrier for O-O bond formation and promote the adsorption of OOH intermediates, a rate-limiting step in OER.[24,25] Similarly, Fe-based HS catalysts exhibit enhanced OER activity due to spin-selective charge transfer, where unpaired electrons mediate efficient hole injection into adsorbed $OH^-$ species.[26] For HER, LS $Ni^{2+}$ in $NiSe_2$ exhibits excellent activity due to reduced Pauli repulsion between filled $d$-orbitals and H $1s$ orbital, weakening $H^*$ binding and enabling rapid desorption of $H_2$.[27]

More recently, our work has demonstrated that spin states in TM-based catalysts can be categorized into *active* and *inert* spin states.[20] Only the active spin state, characterized by unpaired electrons in symmetry-adapted orbitals (such as, out-of-plane $d_{z^2}$ orbital), can significantly modulate hydrogen adsorption strength through symmetry matching with the H-$1s$ orbital. In contrast, the inert spin states with in-plane unpaired electron (such as, $d_{x^2-y^2}$ orbital), exhibit negligible contributions due to symmetry mismatch.[20] However, this simplified binary classification overlooks the complexity of spin-polarized systems. For example, Mn or Fe based catalyst on sulfur-rich supports (such as $MoS_2$) may exhibit distinct spin-dependent behaviors due to stronger metal-support charge transfer.[28–30] Furthermore, while the qualitative distinction between active and inert states is established, the quantitative correlation between the HER activity and active spin states remains unresolved. Addressing these gaps requires establishing correlations between adsorption energetics and spin states, coupled with multi-component interactions within catalyst-support systems, thereby advancing spin state engineering for enhanced HER performance.

Here, we address these challenges by developing a quantitative activity-spin state descriptor for HER, validated across a range of earth-abundant SACs. Specifically, we propose an 'activity index ($\eta$)' as a descriptor, which account for the electronic structure modulation induced by both the intrinsic active site and its local coordination environment, to assess the activity of the spin states for HER. Taking 1T-phase ZrSe$_2$ anchoring TM atoms (TM=Sc to Ni) as models, a linear dependence is found between Gibbs free energy ($\Delta G_H$) and the $\eta$ value, with $\Delta G_H$ declining as $\eta$ decreases. This correlation indicates that the $\eta$ can serve as the effective descriptor for the active spin state in HER on TM-based catalysts. Our work provides fundamental insights into the activity-spin state relationship, offering new avenues for HER catalyst design.

Firstly, we investigate the geometric structure and electronic properties of the 1T-phase ZrSe$_2$. As shown in Fig. 1 (a), the Zr atomic layer is sandwiched between two Se atomic layers, which are staggered relative to each another, forming an octahedral coordination structure. The optimized lattice constants are a = b = 3.793 Å, and the Zr-Se bond length is d = 2.707 Å, which are in agreement with previous results.[31] Here, we employ an anchoring strategy to obtain the SACs. As illustrated in Fig. 1 (b), we select the position directly above the Zr atom as the anchoring site for TM atoms, including Sc, Ti, V, Cr, Mn, Fe, Co and Ni from the fourth period. Table S1 lists the lattice constant of the eight ZrSe$_2$-TM systems, which remain largely unchanged. All binding energies (Table S1) are negative, indicating that these systems are energetically stable.

The electronic properties of ZrSe$_2$-TM systems are fundamentally altered by the introduction of the anchored TM atoms. As illustrated in Fig. 1 (c), the calculated band structure and density of states (DOS) reveal that pristine ZrSe$_2$ is a non-magnetic semiconductor with a band gap of 0.497eV. However, the anchored TM atoms induce a semiconductor-to-metal transition in ZrSe$_2$-TM systems — except for ZrSe$_2$-Ni — as evidenced by the total DOS of these systems in Fig. S1. Furthermore, ZrSe$_2$-Ni remains non-magnetic, while the other systems exhibit magnetic behavior, with their total magnetic moments listed in Table S1.

The SACs with tunable electronic properties upon the same anchoring strategy exhibit interesting catalytic behavior. To evaluate the HER activity of ZrSe$_2$-TM systems, we calculate the

Gibbs free energy ($\Delta G_H$) of hydrogen adsorption by positioning H atoms at the top site of the anchored TM atoms. The optimized configurations are shown in Fig. S2. As illustrated in Fig. 2 (a), the $\Delta G_H$ exhibits a two-segment increasing trend as the anchored TM atom varies from Sc to Ni. Specifically, when the TM atom transitions from Sc to Cr, the $\Delta G_H$ increases from -0.10 to 2.68 eV. In the second segment, when the TM atom varies from Mn to Ni, the $\Delta G_H$ increases from -0.04 to 1.06 eV. Notably, ZrSe$_2$-Mn exhibits the best HER activity with a $\Delta G_H$ of 0.04 eV, which is closer to the thermoneutral ideal value (0 eV) than even Pt ($\Delta G_H$ = -0.09 eV).

The $d$-band center ($\varepsilon_d$) is commonly used to explain the activity of metal and metal-based catalysts, where a higher position of $\varepsilon_d$ typically correlates with stronger adsorption strength.[32,33] For the ZrSe$_2$-TM systems studied here, however, the $\varepsilon_d$ fails to fully account for the observed $\Delta G_H$ variation. As illustrated in Fig. 2 (b), the $\varepsilon_d$ exhibits a monotonic decrease (except for the ZrSe$_2$-Ni semiconductor), which contrasts sharply with the two-segment increasing trend of $\Delta G_H$. Actually, the decrease of $\varepsilon_d$ is due to electron count changes: from ZrSe$_2$-Sc to ZrSe$_2$-Ni, the valence electron count of the anchored TM atoms increases while the transfer electron (Table S1) continuously decreases. Consequently, the increased electron occupation (Fig. 2 (c)) leads to more $d$-orbitals being filled below the Fermi level (E$_F$), thus lowering the $\varepsilon_d$. As a special case, ZrSe$_2$-Ni maintains a relatively high $\varepsilon_d$ due to upward $d$-orbital shifting induced by the semiconductor bandgap. The observed decoupling between the $\varepsilon_d$ and HER activity indicates that the underlying mechanism beyond conventional $d$-band theory governs the catalytic performance in these systems, requiring further investigation.

Interestingly, a surprising correlation between the magnetic moments of anchored TM atoms and their corresponding $\Delta G_H$ is observed in the ZrSe$_2$-TM systems. As shown in Fig. 2 (d), the magnetic moments also exhibit an evident two-segment trend: they increase from Sc to Cr, then decreases from Mn to Ni. This trend strongly suggests that the hydrogen adsorption strength is primarily governed by spin-polarized electronic states, rather than the collective $d$-band effects. Crucially, the localized states near E$_F$, predominantly originating from the anchored TM atoms (Fig. S1), mediate the observed magnetism and likely drive this spin-dependent adsorption mechanism.

In fact, the $\varepsilon_d$ parameter fundamentally accounts for all $d$-orbitals, thus all possible adsorbate-$d$-orbital interactions, but only symmetry-allowed couplings contribute under the selection rules of chemical bonding. At the atop adsorption site of TM atom, the $d_{z^2}$ orbital exhibits symmetry compatibility with the H-1s orbital, forming a σ bond, as shown in Fig. 2 (e). Orbital-projected density of states (PDOS) analysis (Fig. 3) reveals that the localized states (dashed box region) predominantly originate from the $d_{z^2}$ orbital, identifying these as real active states. This is further corroborated by pronounced orbital overlap between the TM-$d_{z^2}$ and H-1s orbitals near $E_F$, as evidenced by wavefunction projection analysis (Figs. 4 and S3). We therefore define these spin-polarized $d_{z^2}$ orbitals as active spin states, whose strong hybridization with H-1s governs hydrogen adsorption strength. Conversely, in non-spin-polarized ZrSe$_2$-Ni system, the $d_{z^2}$ orbital is inert state that show negligible participation in H adsorption.

As illustrated in Fig. 3, the active spin states in ZrSe$_2$-TM systems evolve through three distinct configurations governed by $d$-orbital filling: i) In ZrSe$_2$-Sc system, the single $d$ electron occupies the spin-up $d$-orbital, forming a high-spin state consistent with Hund's rule; ii) For ZrSe$_2$-Ti to ZrSe$_2$-V, $d$ electron partial filling of both spin channels creates intermediate-spin configurations, with ZrSe$_2$-Cr system recovering a high-spin state when spin-up orbitals become fully occupied by $d$ electrons; iii) Beyond half-filling (ZrSe$_2$-Mn to ZrSe$_2$-Ni), progressive spin-down occupation reduces polarization, culminating in ZrSe2-Ni's non-magnetic state with balanced spin-up/down populations. This evolution of spin states reflects the competition between exchange splitting and Coulomb repulsion --- while high-spin states dominate when exchange splitting is large, spin degeneracy emerges upon complete $d$-shell filling. Crucially, the $d_{z^2}$ orbital's spin polarization directly modulates its hybridization strength with H-1s (Figs. 4 and S3), confirming active spin states as the primary determinant of hydrogen adsorption energetics.

Based on the above analysis, as illustrated in Figure 5 (a) (right panel), we propose two idealized types of active spin states. i) Unpaired electron occupation of the $d_{z^2}$ orbital: When only one electron occupies the $d_{z^2}$ orbital, it preferentially occupies the spin-up state (such as ZrSe$_2$-Sc). This unpaired electron actively participates in the reaction, consistent with previous studies.[20] ii) Partial-fully occupied $d_{z^2}$ orbital: When the spin-up state of the $d_{z^2}$ orbital is fully occupied, additional electrons populate the spin-down orbital. Under spin-polarized conditions, the spin-down electron serves as the active spin state, while the spin-up electrons are non-bonding (such as ZrSe$_2$-Mn). Interestingly, for the ZrSe$_2$-TM system, we find that with increased number of valence electrons, the first type of active spin state can evolve into a low-spin state with half spin-up and half spin-down occupancy (such as ZrSe$_2$-Ti), representing a special case of active spin state. In contrast, the second type of active spin state becomes inert after depolarization (such as ZrSe$_2$-Ni). This is verified by the statistical analysis of electron populations in active states (Fig. S4): except for the inert state of ZrSe$_2$-Ni, which has two electrons, all other active spin states exhibit approximately one electron. Regardless of the type of active spin state, only one electron in the $d_{z^2}$ orbital forms a bond with the H 1s orbital, resulting in a fully occupied bonding state and an unoccupied antibonding state. According to molecular orbital theory, this leads to a stronger bonding interaction between the anchored TM atom and hydrogen, thereby enhancing hydrogen adsorption.

In contrast, the $d_{z^2}$ orbital of ZrSe$_2$-Ni, representing an inert state, exhibits spin degeneracy and is relatively far from E$_F$ (Fig. 3). The number of $d_{z^2}$ electrons in the active spin state or inert state is displayed in Fig. S4, showing about 2 for ZrSe$_2$-Ni but 1 for others. This means that the $d_{z^2}$ orbital is fully occupied and its depolarization originates from the second type. As illustrated in Fig. 5 (a) (left panel), the interaction between two electrons in $d_{z^2}$ orbital with the H-1s orbital results in two electrons occupying the bonding orbital and one electron occupies the antibonding orbital. This partial occupancy of antibonding state introduces energetic destabilization of the Ni-H bond, elevates total electronic energy of the system and weakens H adsorption strength. As illustrated in Fig. 4 (d), for ZrSe$_2$-Ni, the interaction between the inert state and H-1s orbital is negligible. The result of -COHP reveals antibonding Ni-H interactions, leading to weak adsorption of the H atom.

Two types of active spin state require refinement in real systems due to substrate-mediated orbital modulation and coordination field effects. Based on the maximum overlap principle of molecular orbital theory, the modified $d_{z^2}$ orbital wavefunction of active spin states (Fig. S5) due to *d*-orbital hybridization and substrate interactions, reduces spatial overlap with H-1s orbitals, thereby reducing the hydrogen adsorption strength. We quantify this effect through the ***active spin state proportion*** (*P*), defined as the weight of active spin states within localized states (Fig. 3). As shown in Fig. 5 (b), *P* decreases across two transition sequences: Sc→Cr (from 0.97 to 0.28) and Mn→Co (from 0.97 to 0.31), directly correlating with weakened interaction between active spin states and H-1s orbital, align with the observed ΔG$_H$ trend. Furthermore, we introduce the **state broadening range** (*R*), defined as the energy span of active spin states, to assess the interactions between $d_{z^2}$ and other orbitals. Figure 5 (c) demonstrates the increased R values along Sc→Cr (from 0.40 to 1.80 eV) and Fe→Ni (from 0.27 to 1.50 eV) sequences, reflecting enhanced delocalization of active spin states. This dual mechanism—reduced hybridization and increased delocalization—synergistically weakens overlap between active spin states and the H-1s orbital, ultimately degrading hydrogen adsorption capability.

Frontier orbital theory further dictates that only the states near the E$_F$ contribute to catalytic activity.[34] Consequently, the $d_{z^2}$ orbitals far from the E$_F$ possess negligible influence on the activity, and are thus excluded from the active spin states. This indicates that the distance between the active spin states and E$_F$ plays a critical role in determining the catalytic activity. Therefore, we define the **active spin state center**, $\varepsilon_{act}$, where $\varepsilon_{act}$ is the integration center of the active spin state. As illustrated in Fig. 5 (d), $\varepsilon_{act}$ shifts downward as the anchored atom varies from Ti to Cr (-0.58 to -1.43 eV) and from Mn to Ni (-0.65 to -1.94 eV), moving the active spin states away from the E$_F$, and aligning with the increase in ΔG$_H$. This trend aligns with the frontier orbital theory, where states closer to E$_F$ exhibit higher reactivity.

However, none of the three individual descriptors can universally account for the catalytic behavior of the studied systems, as their catalytic activity arises from their synergistic interaction among multiple factors. To address this, we propose a unified descriptor — the logarithmic activity index (*η*) — defined as:

$$\eta = \log(|\varepsilon_{act}| \cdot \frac{R}{P})$$

where $\varepsilon_{act}$ represents the energy difference between E$_F$ and integration center of the active spin state, *R* denotes the broadening range of active spin states, and *P* is the proportion of active spin state in localized state.

The activity index *η* demonstrates robust predictive capability for the HER performance across all ZrSe$_2$-TM systems. As illustrated in Fig. 5 (e), when the anchored atom varies from Sc to V, the linear correlation between ΔG$_H$ and *η* is described by: ΔG$_H$ = 0.3712+1.0415*η*. Similarly, for the anchored TM atoms ranging from Mn to Ni, the relationship is given by: ΔG$_H$ = 0.5507+1.0435*η*.

The corresponding coefficients of determination ($R^2$) are 0.9920 and 0.9998, respectively, underscoring the strong linear correlation between $\Delta G_H$ and $\eta$. The nearly identical slopes of the two fits further validate the robustness and transferability of $\eta$ in quantifying the influence of active spin states. According to the Sabatier principle, optimal catalytic performance is achieved when the binding strength between the catalyst and reaction intermediates is neither too strong nor too weak.[35–37] As shown in Fig. 5 (e), a smaller $\eta$ value corresponds to a $\Delta G_H$ closer to thermoneutrality, indicating enhanced HER activity. Among all the TM-based SACs, $ZrSe_2$-Mn has the optimal $\eta$ (-0.56), which exhibits the best HER activity with a $\Delta G_H$ of 0.04 eV. This makes $\eta$ an effective and convenient screening parameter for identifying catalysts with superior intrinsic activity. Thus, $\eta$ serves as a reliable and generalizable descriptor for capturing variations in hydrogen adsorption energetics. It not only provides fundamental insights into HER mechanisms but also enables the rational design of spin-state-engineered electrocatalysts.

In summary, we have investigated HER catalysis by different spin states of TM SACs supported on $ZrSe_2$ using DFT calculations. Two types of active spin states are pinpointed. Considered the electronic structure modulation induced by both the intrinsic active site and its local coordination environment, we propose a descriptor— activity index ($\eta$)— to assess the activity of the spin states for HER. We demonstrate a nearly linear relationship between the $\Delta G_H$ and the $\eta$ value, which fits a relation of $\Delta G_H = 0.3712+1.0415\eta$ ($\Delta G_H = 0.5507+1.0435\eta$) with the determination coefficient $R^2$ as 0.9920 (0.9998). This thus suggests that activity index $\eta$ is a promising HER catalytic descriptor for TM-based catalysts. Our findings provide important insight into the relationship between HER activity and spin, offering new designing strategies for TM-based catalysts.

**SUPPLEMENTARY MATERIALS**

See the supplementary material for the computational methods and additional results.

**ACKNOWLEDGEMENTS**

This work was supported by the National Natural Science Foundation of China (Grants No.

**DATA AVAILABILITY**

The data that support the findings of this study are available within the article and its supplementary material.

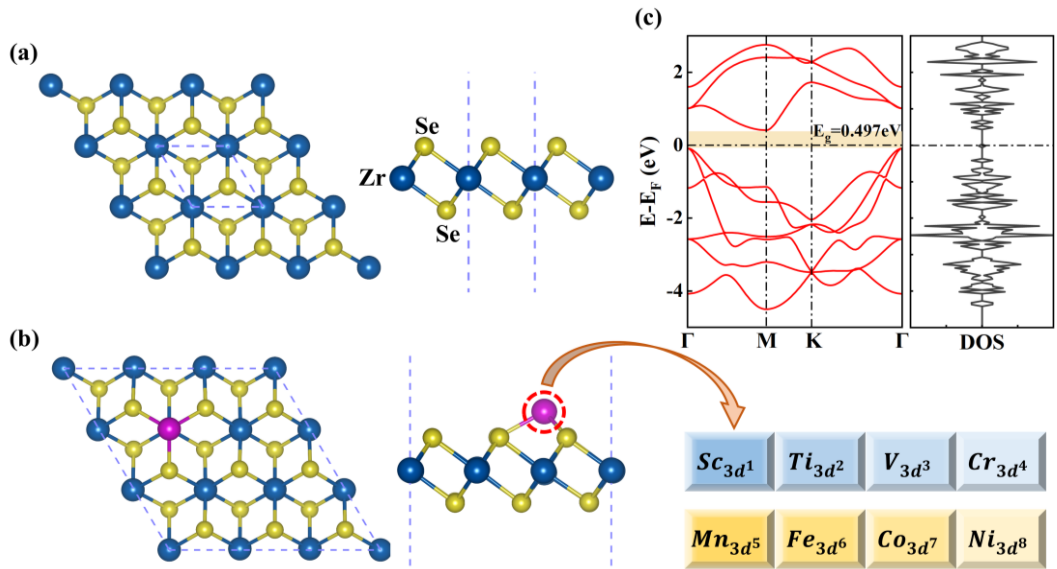

FIG. 1. **The structure, electronic properties and anchoring atoms of 1T-ZrSe$_2$.** (a) The top and side views and (c) electronic band structure and density of state of 1T-ZrSe$_2$. The Fermi level is set to zero. (b) The top and side views of 1T-ZrSe$_2$-TM with primitive unit cell shown by the dashed line. On the right are electron configurations of the eight fourth period metal elements anchored on ZrSe$_2$.

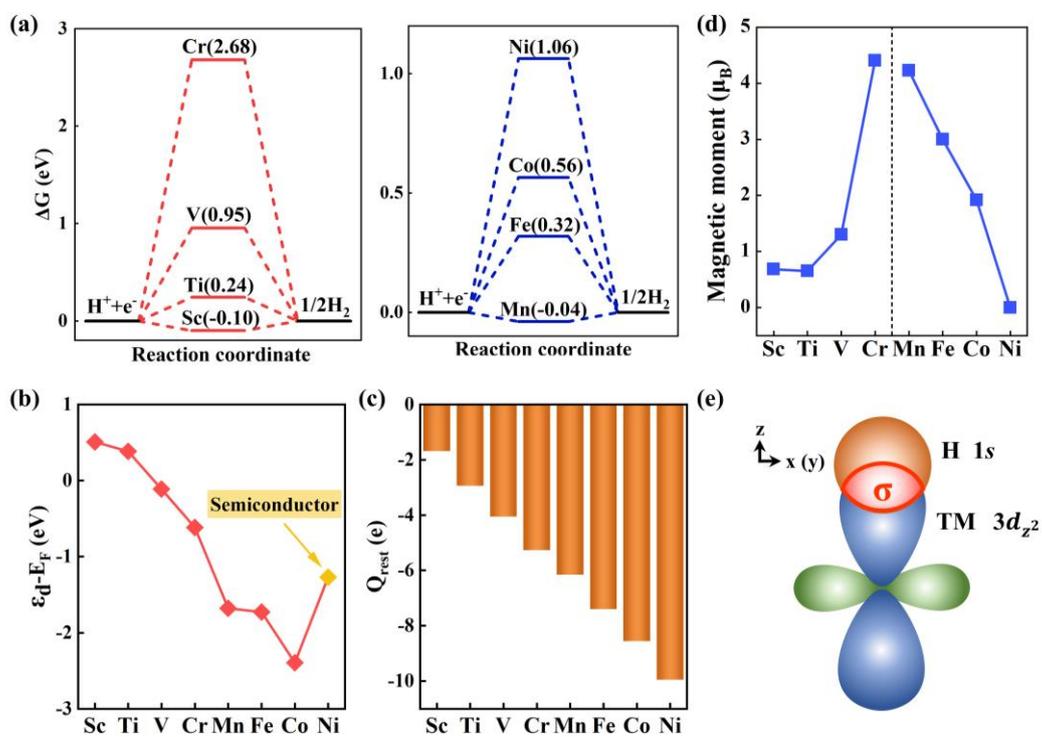

FIG. 2. Calculated Gibbs free energy diagram for HER on TM site of (a1) ZrSe$_2$-TM (TM=Sc, Ti, V, Cr) and (a2) ZrSe$_2$-TM (TM=Mn, Fe, Co, Ni). The (b) d-band center, (c) residual electron and (d) magnetic moment of TM atom of ZrSe$_2$-TM. (e) Diagram of TM-H orbit interaction.

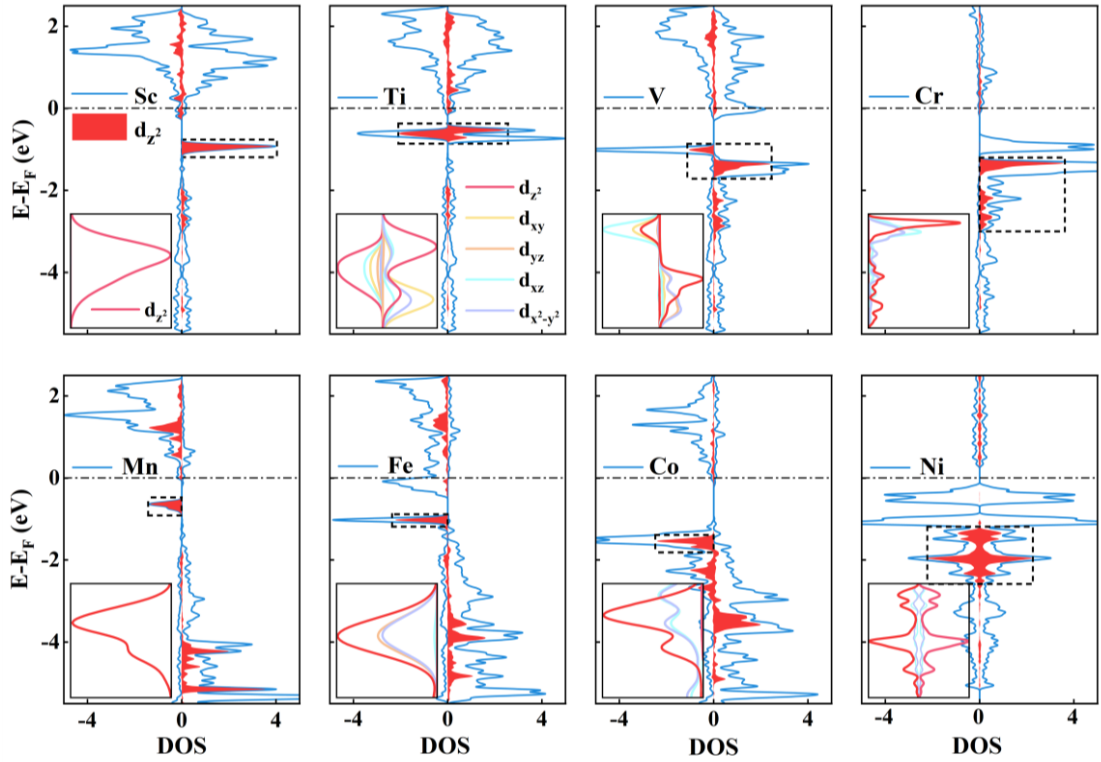

FIG. 3. The PDOS of ZrSe$_2$-TM (TM =Sc, Ti, V, Cr, Mn, Fe, Co, Ni) and local magnification of the active spin states. The local magnification in the lower left corner is the local enlargement in the d orbital of five directions of the state boxed by dashed line. The Fermi levels are set to zero.

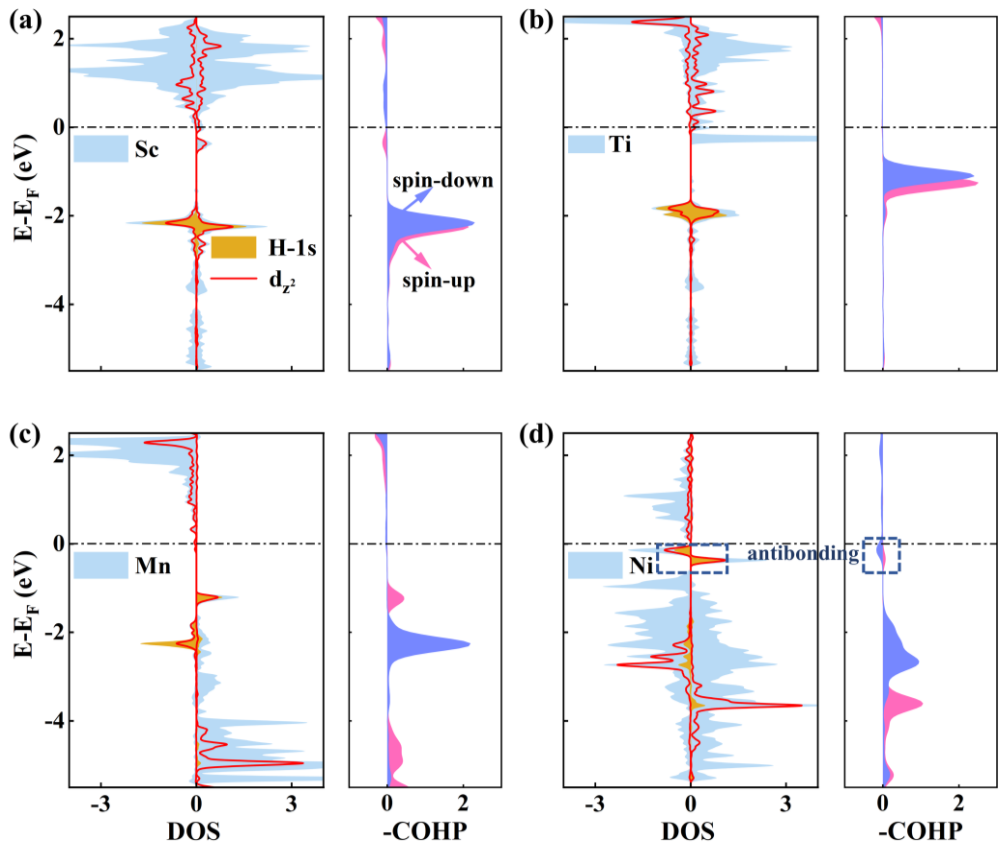

FIG. 4. The PDOS and -COHP of $ZrSe_2$-TM-H* (TM = Sc, Ti, Mn, Ni). The Fermi levels are set to zero.

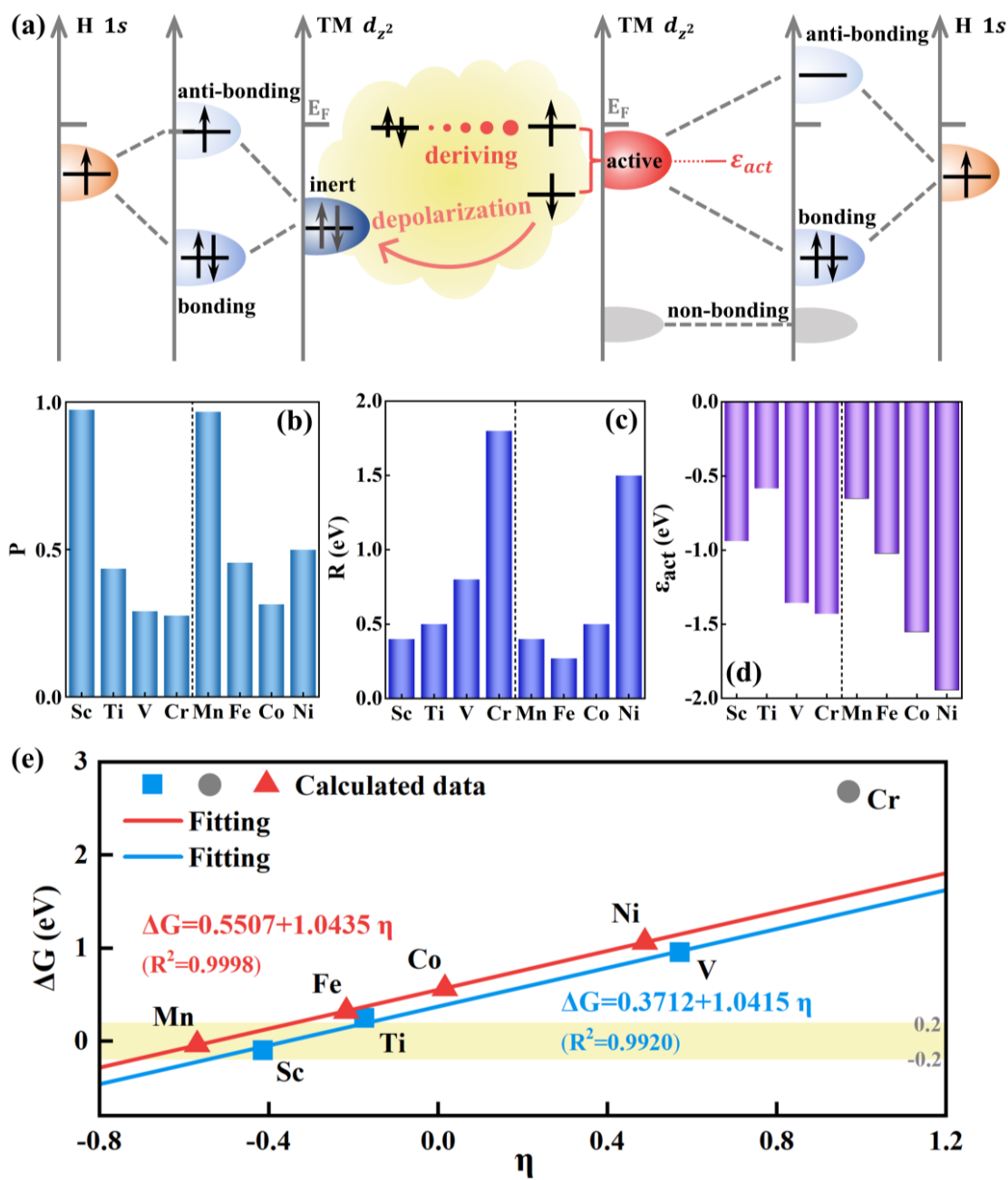

FIG. 5. **The HER mechanism of active spin states.** (a) Diagram of the interaction between the active spin state (inert state) and H-1s, two idealized types of the active spin states and relationship between the active spin state and inert state. The (b) *proportion* P, (c) *broadening range* R and (d) *center* $\varepsilon_{act}$ of active spin states. (e) Linear relation between the logarithmic activity index $\eta$ and Gibbs free energy $\Delta G$.